\documentclass{jetpl}
\usepackage{amsfonts}
\usepackage{amssymb}
\usepackage{psfig}
\usepackage{mathrsfs}
\twocolumn
\lat
\newcommand{\beq}{\begin{equation}}
\newcommand{\eeq}{\end{equation}}
\newcommand{\beqn}{\begin{eqnarray}}
\newcommand{\eeqn}{\end{eqnarray}}
\newcommand{\bea}[1]{\beq\begin{array}{#1}}
\newcommand{\eea}{\end{array}\eeq}

\newcommand{\sign}{\mathop{\rm sign}}

\newcommand{\MeV}{{\text{MeV}}}
\newcommand{\fm}{{\text{fm}}}
\newcommand{\CUT}{50}
\newcommand{\zvlZ}{3.7(5)}
\newcommand{\zvlO}{0.38(5)}
\newcommand{\zvlChi}{3}
\newcommand{\lvlZ}{1.9(2)}
\newcommand{\lvlO}{2.26(3)}
\newcommand{\lvlChi}{2}
\newcommand{\zaFourChi}{0.95}
\newcommand{\zaThreeChi}{0.38}
\newcommand{\zaTwoChi}{0.85}
\newcommand{\laFourChi}{2.3}
\newcommand{\laThreeChi}{3.4}
\newcommand{\laTwoChi}{4.9}

\title{Evidence for fine tuning of fermionic modes in lattice gluodynamics}
\rtitle{Fine tuning of fermionic modes \ldots}
\sodtitle{Evidence for fine tuning of fermionic modes in lattice gluodynamics}

\author{F.\,V.\,Gubarev$^*$, S.\,M.\,Morozov$^*$, M.\,I.\,Polikarpov$^*$, V.\,I.\,Zakharov$^+$}
\rauthor{F.\,V.\,Gubarev, S.\,M.\,Morozov, M.\,I.\,Polikarpov, V.\,I.\,Zakharov}
\sodauthor{Gubarev, Morozov, Polikarpov, Zakharov}

\address{Institute of Theoretical and  Experimental Physics, B.~Cheremushkinskaya, 25, Moscow, 117259, Russia \\ ~ \\
         Max-Planck Institut f\"ur Physik, F\"ohringer Ring 6, 80805, M\"unchen, Germany}

\dates{18 July 2005}{*}


\abstract{ We consider properties of zero and near-zero fermionic modes
in lattice gluodynamics. The modes are known to be sensitive
to the topology of the underlying gluonic fields in the quantum vacuum state of
the gluodynamics. We find evidence that these modes
are fine tuned, that is exhibit sensitivity to both physical
(one can say, hadronic) scale and
to the ultraviolet cut off. Namely, the density of the
states is in physical units while the localization volume of the modes tends to zero in physical
units with the lattice spacing tending to zero. We discuss briefly
possible theoretical implications and also  include some general, review-type remarks. }

\begin{document}
\maketitle
%=============================================================================
\section{Two scales of QCD}
In this note we will consider properties of vacuum fluctuations
within lattice formulation of Yang Mills theories. For simplicity,
we concentrate on the case of pure gluodynamics, with no dynamical
fermions. What is specific for the lattice formulation
(see, e.g., \cite{creutz}) is that
it is a field theory in Euclidean space-time.
The action reads:
\begin{equation}\label{action}
S~=~{1\over 4 g^2}\int d^4x G_{\mu\nu}^a(x)G_{\mu\nu}^a(x)~~,
\end{equation}
where $  G_{\mu\nu}^a(x)$ is the non-Abelian field strength tensor
and $a$ is the color index, g is the coupling. We will consider actually the SU(2) case,
$a=1,2,3$. Using action (\ref{action}) one generates vacuum field configurations,
$\{A_{\mu}^a\}$, where $A_{\mu}^a$ is the gauge potential
and performs further measurements on these fields.

As any renormalizable theory, quantum gluodynamics  exhibits two
scales, infrared and ultraviolet. Moreover, the ultraviolet cut off is
introduced explicitly, through a finite lattice spacing $a$. The
infrared scale, $\Lambda_{QCD}$, on the other hand  is emerging
dynamically:
\begin{equation}
\Lambda_{QCD}^2~\approx~{1\over a^2}\cdot \exp{(~-b_0/g^2(a))}~~,
\end{equation}
where $b_0$ is a constant, $g(a)$ is the bare coupling constant normalized at the
lattice spacing, $\Lambda_{QCD}$ characterizes the scale where the running coupling
is of order unit. As $a\to 0$, the bare coupling squared, $g^2(a)$ tends to zero as an inverse
log of $(a\cdot\Lambda_{QCD})$.
If one changes the lattice spacing
$a$ and modifies $g(a)$ according to the rules of the renormgroup
the scale $\Lambda_{QCD}$ does not depend on $a$.

Lattice simulations allow to study directly vacuum fluctuations  and
the both scales, $\Lambda_{QCD}$ and  $1/a$ get manifested
in the vacuum fluctuations.
In particular, the zero-point fluctuations are sensitive to the
ultraviolet cut off.
One can measure them, for example, by studying the gluon condensate, or
the vacuum expectation value of the gluonic field strength tensor squared:
\begin{equation}
\label{gmn}
\langle 0|~{g^2\over 32\pi^2} (G_{\mu\nu}^a)^2~|0\rangle~\approx~{const\over a^4}\{1+\Sigma_k a_kg^{2k}(a)\}
\end{equation}
where  $a_k$ are coefficients of the
perturbative series.
The matrix element (\ref{gmn}) is divergent as the fourth power of the ultraviolet cut off.
This is the well known divergence of the density
of the vacuum energy in field theory, which arises due to the zero-point
fluctuations. High gluonic frequencies dominate
the matrix element (\ref{gmn}) because of the phase space associated with these
fluctuations. One can say that zero-point fluctuations represent
an example of entropy dominated fluctuations.

Situation looks absolutely different if one considers topological charge.
The density of the topological charge is given by:
\begin{equation}
Q(x)~=~{g^2\over 32\pi^2}\epsilon_{\mu\nu\rho\sigma}G_{\mu\nu}^a(x)G_{\rho\sigma}^a(x)~~,
\end{equation}
where $\epsilon_{\mu\nu\rho\sigma}$ is the totally antisymmetric
tensor. One can readily show that $Q(x)$ is a full derivative. As a
result, all perturbative fluctuations do not contribute to the
topological charge. On the other hand, probability to find a
non-perturbative fluctuation of size $\rho$ is suppressed for small
$\rho$ as $\exp(-const/g^2(\rho))$ where $g(\rho)$ is the running
constant. This factor grows fast with growing $\rho$ and is somehow
smoothened out at $\rho \sim\Lambda_{QCD}^{-1}$ where the running of
the coupling cannot be calculated reliably. As a result, topologically
nontrivial fluctuations have typical size of order $\Lambda_{QCD}^{-1}$
and are absolutely negligible on the scale of $a$. In particular,
instantons are topologically non-trivial and all the factors mentioned
above are known explicitly. Instantons represent fluctuations whose
probability is determined primarily by action, not entropy
(for a review of instantons, in the context of the lattice measurements see, e.g.,
\cite{instanton-reviews}).

A common viewpoint is that confinement is due to
vacuum fluctuations on the scale $\Lambda_{QCD}$,
i.e. is determined by soft, semiclassical fields.
The expectations can be confronted with lattice measurements
on the fully quantum, vacuum  state of gluodynamics.
And most recently there has been emerging evidence \cite{fine-tune} that
the confining fields are actually of a third type, exhibiting both
infrared and ultraviolet scales. Both action and entropy are very large but
balance each other almost exactly. One can call them fine tuned
fluctuations, for a review see \cite{vz}.
\section{Fine tuning}
Imagine that a relativistic system has size $r_0$. Then, typical momenta should be of order
$p\sim1/r_0$. Respectively, one expects that typical masses are of order $m^2\sim 1/r_0^2$.
If, on the other hand, observed masses are much smaller than $r_0^{-2}$,
one calls such a case fine tuning.

The notion of fine tuning has been discussed most thoroughly
in connection with the Higgs physics. The problem here is that to fulfill
its role as a part of a renormalizable theory of weak interactions
the Higgs particle should have mass of order $100~\mathrm{GeV}$, like intermediate bosons
of weak interactions. On the other hand, if Higgses are point-like particles
down to scale $r_0$, then
radiative  corrections to the Higgs mass are of order
\begin{equation}
\Delta m_H^2~\sim~\alpha r_0^{-2}~~,
\end{equation}
where $\alpha$ is the electromagnetic coupling. If $r_0$ is, say,
of order of inverse Planck mass, fine tuning of the radiative correction
and of the bare mass is required.

To resolve the puzzle of fine tuning, if it is observed experimentally, one usually
invokes hidden symmetry. In the Higgs case, for example, one of the favorite
candidates for such a symmetry now is supersymmetry.

In case of Yang-Mills theories, one usually does not expect to confront
the problem of fine tuning. However, recently it was found that the
field configurations which are responsible for the confinement appear
fine tuned. In particular, there is an ample evidence that so called
central vortices, for review see, e.g., \cite{vortices}, are
responsible for the confinement. The central vortices represent closed
two-dimensional surfaces whose total area scales  in physical units:
\begin{equation}
\label{ir}
A_{tot}~\approx~ 24  \,\, V_{tot} \,\,\, \mathrm{fm}^{-2}
\end{equation}
where $V_{tot}$ is the total volume of the lattice.
On the other hand, the non-Abelian action associated with the surfaces is
ultraviolet divergent \cite{fine-tune}:
\begin{equation}\label{uv}
S_{tot}~\approx~0.55~{A_{tot}\over a^2}~~.
\end{equation}
Combining observations (\ref{ir}) and (\ref{uv}) one concludes that
the suppression of the fluctuations due to the action (\ref{uv}) is to be nearly compensated by
enhancement due to the entropy. That is, the fluctuations are fine tuned.

The observations above are pure empirical. Theoretically, it is natural to speculate again
that there is a hidden symmetry which ensures the observed fine tuning.
Moreover, the only symmetry which can come into consideration is the
conjectured duality between Yang-Mills
theories and string theories \cite{maldacena}. However, such a connection is pure
speculative at the moment \cite{vz}. We mention this possibility just to emphasize
that further studies of the  fine tuning in Yang-Mills theories are of great interest.
In this paper, we address probable manifestation  of the fine tuning
in fermionic zero modes.

\section{Witten-Veneziano, Banks-Casher relations}

 We will study properties of low-lying modes of the Dirac operator,
\begin{equation}
D_{\mu}\gamma_{\mu} \,\, \psi_n ~=~ \lambda_n\psi_n\,,
\end{equation}
where $\gamma_{\mu}$ are the Dirac matrices and the covariant
derivative $D_{\mu}$ is constructed on the gauge potential $A_{\mu}^a$
generated as a vacuum field configuration. The Dirac equation is solved
numerically. Moreover,  one generates many configurations and studies
the properties of the modes with low values of $\lambda_n$. Note that
both $\lambda_n$ and the volume occupied by the n-th wave function
$\psi_n$ are gauge invariant quantities.

There exists a rich literature on the low-lying Dirac eigenmodes (LDEs).
Most commonly, one uses the instanton model of the vacuum, for review and
further references see, e.g.,
\cite{instanton-reviews}.
The instantons are localized solutions with non-trivial topological charge
and there are exact zero fermionic modes associated with them:
\begin{equation}\label{zero}
n_{+} ~-~ n_{-} ~=~ N_f\cdot Q_{top}
\end{equation}
where $n_{\pm}$ are the number of zero modes with positive or negative chirality,
$N_f$ is the number of quark flavors and the topological charge $Q_{top}=1$ for an instanton.

In the physical vacuum, the instanton solutions are distorted.
Indeed, there are neighboring instantons (or anti-instantons) and the instanton
fields are modified because of that. The corresponding fermionic
zero modes are becoming near-zero modes.

There are some generic features which are predicted to survive the
modifications. First, consider exact zero modes, for a given lattice
volume $V_{tot}$. The exact zero modes are related to the total
topological charge $Q_{top}$. On average, $Q_{top}=0$. The value of
$Q^2_{top}$ fluctuates, however. The instanton-like picture presumes
that there are lumps of topological  charge close to $Q_{top}=\pm 1$
occupying sub-volumes of order $\Lambda_{QCD}^{-4}$. There are no
fluctuations of the topological charge on the scale of the lattice
spacing $a$. Indeed, instantons are topological excitations with
smallest action possible and the probability to find instanton of size
of order $a$ is proportional to a high power of $a$. Quantitatively,
the strength of the topological fluctuations is related to the
$\eta^{'}$ mass \cite{wittenveneziano}:
\begin{equation}\label{chi}
\chi_t~\equiv~{\langle Q^2_{top}\rangle\over V_{tot}}~\approx~{m_{\eta^{'}}^2f_{\pi}\over 2N_f}
~\approx~~(213~MeV)^4\,,
\end{equation}
where the numerical value is borrowed from Ref \cite{spacings}.

There are also near-zero fermionic modes whose total number is proportional to the total volume
and which are associated, in the zero approximation, with the original lumps of the topological charge,
see above. Since the modes are nearly degenerate, the interaction between instantons
results in the delocalization of these zero modes, for details see, e.g.,  \cite{instanton-reviews}.
Quantitatively, this generic feature of the instantonic vacuum is manifested
through the Banks-Casher relation \cite{banks-casher}:
\begin{equation}\label{rho}
\langle \bar{\psi}\psi\rangle~=~-\pi \rho(\lambda_n\to 0)~~,
\end{equation}
where $\rho(\lambda_n\to 0)$ is the density of the (delocalized) zero modes in the limit of
infinite volume $V_{tot}$. For finite lattice volumes these modes are near-zero.

Thus, Eqs (\ref{chi}) and (\ref{rho}) predict that
the number of exact zero modes is proportional to $\sqrt{V_{tot}}$ while the number
of near-zero modes is proportional to the total volume, $V_{tot}$. Moreover, the instantonic
picture fixes the scaling laws according to which the corresponding coefficients of
proportionality are in physical units and not dependent on the lattice spacing $a$.

\section{Measurements}

We are using the overlap Dirac operator \cite{neuberger}. The advantage
of the overlap operator is that it preserves the chiral symmetry and
allows to study the properties of the Dirac modes from first principles
\cite{luscher}. There were actually many studies of the
Witten-Veneziano and Banks-Casher relations on the lattice. However
they used mostly such versions of the lattice fermions that the
topological charge has intrinsic ultraviolet noise. To avoid this, one
has to modify, smoothen the original gauge fields which fluctuate on
the lattice spacing $a$, for review see, e.g., \cite{teper}. As a
result, no dependence on $a$ could actually be measured. The use of the
overlap fermions allows to study original gauge fields configuration,
see, in particular, \cite{horvath}. Study of dependence on the lattice
spacing $a$ of various observables is one of the main objectives of the
present paper.

More explicitly,  the massless overlap Dirac operator
is given by~\cite{neuberger}:
\beq
\label{overlap}
D = \frac{\rho}{a} \left( 1 + \frac{A}{\sqrt{AA^\dagger}} \right),\,\quad
A = D_W - \frac{\rho}{a},
\end{equation}
where $A$ is the Wilson Dirac operator with negative mass term.
Anti-periodic (periodic) boundary  conditions in time (space) directions
were employed. It turns out that the optimal value of $\rho$
parameter is $1.4$.  Furthermore, we have used the minmax polynomial
approximation~\cite{giusti}
to compute the sign function $\sign(A) = A/\sqrt{AA^\dagger} \equiv \gamma_5 \sign(H)$,
where $H=\gamma_5 A$ is hermitian Wilson Dirac operator.
In order to improve the accuracy and performance about one hundred
lowest eigenmodes of $H$ were projected out.
Note that the eigenvalues of (\ref{overlap}) lies on the circle of radius
$\rho$ centered at $(\rho,0)$ in the complex plane. In order to relate them
with continuous eigenvalues of the Dirac operator the circle was
stereographically projected onto the imaginary axis~\cite{capitani}.

The calculations were performed on two subsets (Table \ref{tab:params}) of
statistically independent $SU(2)$ quenched configurations generated with
standard Wilson action. For the subset A the gauge coupling was chosen is such
a way that the physical volume remains the same, the corresponding lattice
spacings were determined by interpolating the data of Ref.~\cite{spacings}.
The subset B was used to determine the IPR volume dependence  at fixed spacing
$a = 0.1394(8)\,\fm$.

\begin{table}
%\begin{ruledtabular}
\begin{tabular}{p{0.08\textwidth} p{0.1\textwidth} p{0.04\textwidth}
                p{0.04\textwidth} p{0.1\textwidth} p{0.04\textwidth} }
$\beta$ & $a$,fm & $L_s$ & $L_t$ & $V^{phys}, \mathrm{fm}^4$ & $ N_{conf}$ \\ \hline
\multicolumn{6}{c}{Subset A} \\ \hline
2.3493  & 0.1397(15) &  10   & 10    & 3.8(2) & 300 \\
2.3772  & 0.1284(15) &  10   & 14    & 3.8(2) & 91  \\
2.3877  & 0.1242(15) &  10   & 16    & 3.8(2) & 198 \\
2.4071  & 0.1164(15) &  12   & 12    & 3.8(2) & 179 \\
2.4180  & 0.1120(15) &  12   & 14    & 3.8(2) & 149 \\
2.4273  & 0.1083(15) &  12   & 16    & 3.8(2) & 198 \\
2.4500  & 0.0996(22) &  14   & 14    & 3.8(3) & 200 \\
2.5000  & 0.0854(4)  &  16   & 18    & 3.92(7)& 196 \\
\hline
\multicolumn{6}{c}{Subset B} \\ \hline
2.3500  & 0.1394(8)  &  10   & 10    & 3.8(1) & 100 \\
2.3500  & 0.1394(8)  &  10   & 14    & 5.3(1) & 100 \\
2.3500  & 0.1394(8)  &  12   & 12    & 7.8(2) & 100 \\
2.3500  & 0.1394(8)  &  12   & 18    &11.7(2) & 100 \\
2.3500  & 0.1394(8)  &  14   & 14    &14.5(3) & 94  \\
2.3500  & 0.1394(8)  &  14   & 16    &16.5(4) & 42  \\
\end{tabular}
%\end{ruledtabular}
\caption{Simulation parameters.}
\label{tab:params}
\end{table}

\section{Results for topological susceptibility}

\begin{figure}
\centerline{\psfig{file=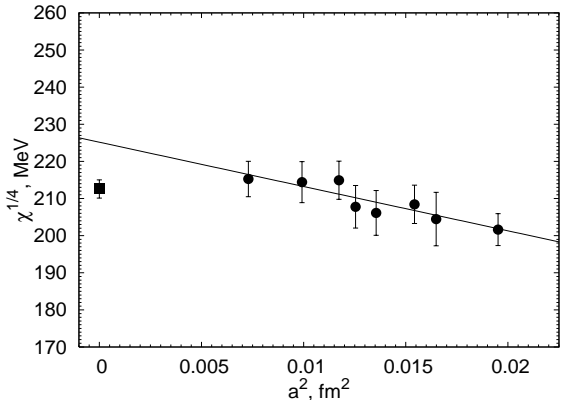,width=0.45\textwidth,silent=,clip=}}
\caption{Scaling of topological susceptibility with lattice spacing; solid curve is the
best fit $\chi^{1/4} = \chi_0^{1/4} + c\, a^2$. Square corresponds to the conventional
value $(213(3)\,\MeV)^4$, ~\cite{spacings}}
\label{fig:chi_vs_a}
\end{figure}

Let us address first the issue of  the scaling of topological susceptibility
$\chi_t = \langle Q^2_{tot}\rangle/V_{tot}$, where the topological charge is defined by the
index of overlap Dirac operator $Q_{top} = n_+ - n_-$ and $n_\pm$ is the number
of exact zero modes with positive (negative)  chirality. The topological
susceptibility calculated on the subset A scales up to the $a^2$ corrections
(Figure~\ref{fig:chi_vs_a}) and equals to $\chi_0 = (225(3)\,\MeV)^4$
which is in good agreement with the conventional value $(213(3)\,\MeV)^4$~\cite{spacings}.

%----------------------------------------------
%----------------------------------------------
\begin{figure}
\centerline{\psfig{file=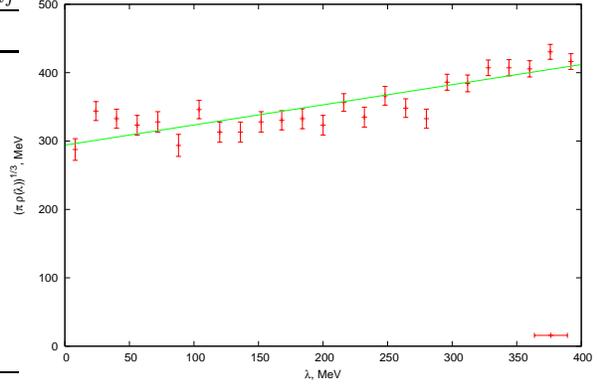,width=0.45\textwidth,silent=,clip=}}
\caption{The spectral density of eigenmodes of the overlap Dirac operator and its interpolation to
$\lambda_n \to 0$ at $\beta=2.5$ on $18\times 16^3$ lattice.
The value of the condensate $\langle\bar\psi\psi\rangle$, Eq.~(\ref{rho}),
is perfectly consistent with results of Ref.~\cite{hands}.}
\label{fig:rho_vs_lambda}
\end{figure}
%----------------------------------------------
\begin{figure}
\centerline{\psfig{file=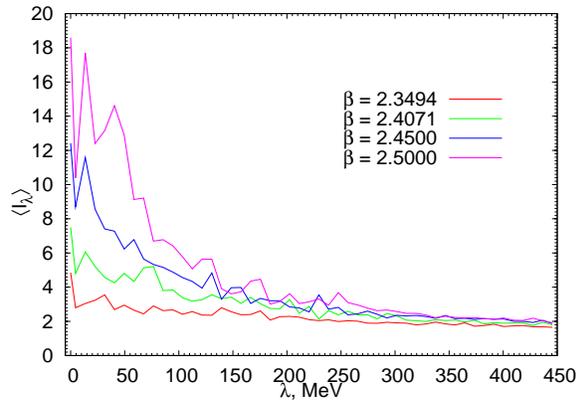,width=0.45\textwidth,silent=,clip=}}
\caption{IPRs for LDEs at various lattice spacings and fixed physical volume.
The ``mobility edge'' $\lambda_{cr} \approx 200\,\MeV$ is clearly seen.}
\label{fig:ipr_histo_a}
\end{figure}
%----------------------------------------------

The observed scaling of the topological susceptibility
confirms that the discretization errors and the finite volume effects
are small. Note that the Wilson gauge action is plagued by lattice dislocations
and one might have speculated that spurious fermionic zero modes
are generated. However, Figure~\ref{fig:chi_vs_a}
confirms that the dislocations are inessential for overlap Dirac operator
which is insensitive to the ultraviolet noise.

Moreover, we have measured the density of near-zero  modes, see Figure~\ref{fig:rho_vs_lambda}.
The value of the condensate $\langle\bar{\psi}\psi\rangle$ determined through
the Banks-Casher relation turns to be close
to the value obtained earlier with the Wilson fermions, see, for instance,\cite{hands}.

\section{Shrinking of the volume occupied by fermionic modes}

A natural measure of the eigenmodes localization is provided by the
inverse participation ratio (IPR) $I_{\lambda}$ which is defined as
follows (for review and applications see for example Ref.~\cite{CM}).
Let $\rho_{\lambda}(x)$ be the normalized bilinear
$$
\rho_\lambda(x) = \psi_\lambda^\dagger(x)\psi_\lambda(x)\,,\qquad
\sum_x \rho_\lambda(x) = 1\,,
$$
where $\psi_{\lambda}(x)$ is the eigenmode of the overlap Dirac operator in the given
gauge field background with virtuality $\lambda$, $D\,\psi_{\lambda} = \lambda\,\psi_{\lambda}$.
Then for any finite volume $V$ the IPR $I_{\lambda}$ is defined by
\beq
\label{eq:ipr_def}
I_\lambda = V \, \sum_x \rho_\lambda^2(x)\,,
\end{equation}
and characterizes the inverse fraction of sites contributing to the support of $\rho_{\lambda}(x)$.
Note that for delocalized modes $\rho_{\lambda}(x) = 1/V$ and hence $I_{\lambda} = 1$,
while an extremely localized mode, $\rho_{\lambda}(x) = \delta_{x,x_0}$,
is characterized by $I_\lambda = V$. Moreover, for eigenmodes localized on fraction $f$ of sites
(so that $\mathrm{sup}~\rho_\lambda = V_f = f ~ V$) we have $I_{\lambda} = V/V_f = 1/f$.
If we allow for a mixture of both localized and extended modes the average value of IPR is given by:
\begin{equation}\label{eq:ipr_mix}
\langle I_{\lambda} \rangle = c_0  + c_1 V / V_f.
\end{equation}

Usually \cite{CM}, one considers localization only in terms of the total volume.
Very recently, there appeared data that the localization volume
can depend on the lattice spacing as well. Namely, such effect was observed first
in Ref. \cite{0410024} for fermionic modes (in a particular version of lattice fermions,
so called Asqtad fermions). Most recently localization properties
of scalar probe particles were investigated \cite{greensite} and, again, strong dependence
on $a$ was observed. Observed dependence on the lattice spacing seems specific
for dynamics of Yang-Mills theories.

The localization properties of LDEs in our measurements are illustrated on
Figure~\ref{fig:ipr_histo_a} where we plot the inverse participation
ratios for modes $0 \le \lambda \le 450 ~\MeV$
at various spacings and fixed physical volume (subset A, Table~\ref{tab:params}).
One can see that there is a critical value,
$\lambda_{cr} \approx 200~\MeV$, above which all the states are in fact delocalized
and their IPRs are lattice spacing independent. However, for small eigenvalues below
$\lambda_{cr}$ the value of IPR grows and one cannot exclude
presence of localized states. The mixture of localized and extended modes could be
quantitatively characterized by the volume dependence of IPR at fixed lattice spacing,
see Eq~(\ref{eq:ipr_mix}).
We have computed the average value
$I_l =  \sum_{\lambda} \langle I_\lambda \rangle /N_l$ of IPRs
for modes $0 < \lambda \le \Lambda  = \CUT ~\MeV$, where $N_l$ is the total number of
modes in this interval, and the analogous quantity $I_0$ for exact zero modes.
Fitting the data to Eq. (\ref{eq:ipr_mix}) we found for exact zero modes
$c_0 = \zvlZ,\, c_1 = \zvlO$ with $\chi/n.d.f = \zvlChi$, and for low-lying
modes $c_0 = \lvlZ,\, c_1 = \lvlO$ with $\chi/n.d.f = \lvlChi$.

\begin{figure}
\centerline{\psfig{file=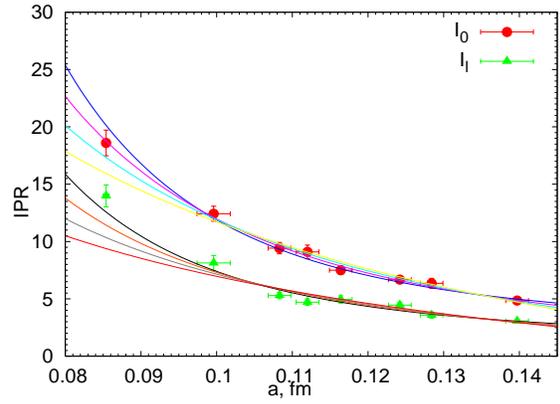,width=0.45\textwidth,silent=}}
\caption{Scaling of $I_0$ and $I_l$ with $a \rightarrow 0$.
Lines correspond to the fitting curves (\ref{volume}) for $d=0,1,2,3$.}
\label{fig:ipr_vs_a}
\end{figure}

Moreover, one can see from Figure~\ref{fig:ipr_histo_a} that
the LDE's localization degree depends non-trivially on
the lattice resolution. To quantitatively investigate this dependence
we have plotted $I_0$ and $I_l$ against the lattice spacing on Figure~\ref{fig:ipr_vs_a}.
It turns out that our results are highly robust with respect to the actual value of
$\Lambda$ as long as $\Lambda \lesssim \CUT~\MeV$ so that we could take $\Lambda = \CUT~\MeV$
to improve the statistics.  It follows from Figure~\ref{fig:ipr_vs_a} that
the inverse participation ratios for both zero and low lying eigenmodes
seem to be divergent in the limit $a\to 0$.
Note that the average values of IPRs are higher than those of Ref.~\cite{0410024}.

%---------------------------------------------------
\begin{figure}
\centerline{\psfig{file=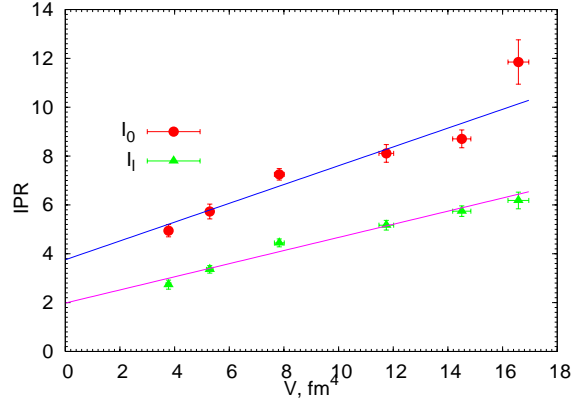,width=0.45\textwidth,silent=}}
\caption{Volume dependence of $I_0$, $I_l$ at fixed lattice spacing.}
\label{fig:ipr_vs_vol}
\end{figure}
%---------------------------------------------------
%---------------------------------------------------

Furthermore, it is straightforward to estimate the dimensionality $d$ of the objects which localize
the low lying eigenmodes. Indeed, at fixed physical volume we have
$$
I_\lambda = V \slash \,V_f = V a^4 \slash \, (V_f a^4) = a^{d-4} \,\, V^{phys} \slash \, V_f^{phys}
$$
and therefore the lattice spacing dependence is given by
\begin{equation}
\label{volume}
\langle I_\lambda \rangle ~=~ b_0 + b_1 \cdot a^{d-4}\,.
\end{equation}
We fitted our data to Eq.~(\ref{volume}) for $d=0,1,2,3$, the fitting curves are shown
on Figure~\ref{fig:ipr_vs_a}. As a matter of fact the IPR data for low lying modes strongly suggest that
the dimensionality of underlying objects is zero, $d=0$; the relevant $\chi^2/n.d.f.$ is $\laFourChi$
which should be compared with its values $\laThreeChi$ ($d=1$) and $\laTwoChi$ ($d=2$).
As far as the exact zero modes are concerned our data favor one dimensional ($d=1$) localization regions;
$\chi^2/n.d.f.$ in this case is $\zaThreeChi$ while it is $\zaFourChi$ for $d=0$ and $\zaTwoChi$ for $d=2$.
The error bars are not small enough, however, to rule out reliably that the exact zero modes
in the limit $a\to 0$ are localized also on point-like objects.

\section{Effect of the removal of the vortices}

Thus, our results indicate fine tuning of the fermionic modes. Namely,
their volume shrinks without affecting the eigenvalues. We have already
mentioned (see Eqs.~(\ref{ir},\ref{uv})) that fine tuning was observed
first for the central vortices. A natural question arises, whether the
fine tuning of the fermionic modes is related to existence of the fine
tuned vortices. A standard way to probe such a relation is the so
called removal of the vortices \cite{forcrand, gattringer}. Namely one
multiplies the original link matrices $U_{\mu}(x)$ by their $Z_2$
projected values:
\begin{equation}\label{modification}
U_{\mu}(x)~\to~U_{\mu}(x)\cdot (Z_2)_{\mu}(x)~~.
\end{equation}
One can show that the modification (\ref{modification}) affects (up to
a gauge transformation) only a 3d fraction of the total 4d volume of
the lattice \cite{volumes}. In other words, this fraction is
proportional to $(a\cdot\Lambda_{QCD})$ and vanishes in the limit of
$a\to 0$. On the other hand, the vortices are removed. For

\begin{figure}
\centerline{\psfig{file=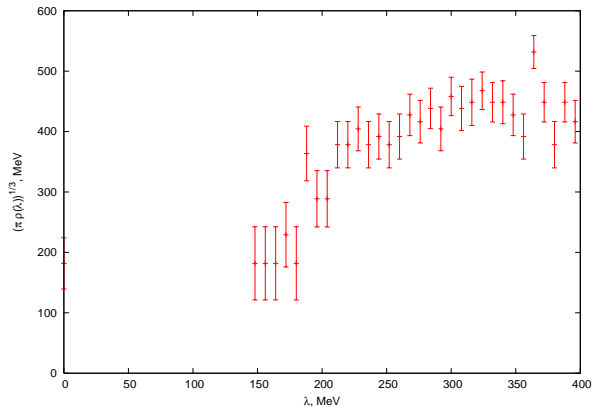,width=0.45\textwidth,silent=,clip=}}
\caption{The spectral density of overlap Dirac operator eigenmodes for vortex removed configurations.
The lattices used are $18\times 16^3$ at  $\beta=2.5$.}
\label{fig:rho_vs_lambda_vortex_removed}
\end{figure}

\begin{figure}
\centerline{\psfig{file=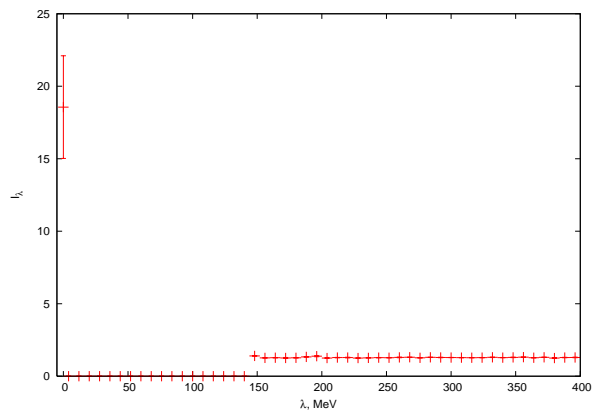,width=0.45\textwidth,silent=,clip=}}
\caption{IPR for LDEs for vortex removed configurations obtained at $\beta=2.5$ on $18\times 16^3$ lattices.}
\label{fig:ipr_vs_lambda_vortex_removed}
\end{figure}

The impact of the change (\ref{modification}) on the properties of LDEs
is illustrated on
Figures~\ref{fig:rho_vs_lambda_vortex_removed},~\ref{fig:ipr_vs_lambda_vortex_removed}.
Namely, all the low-lying modes, with exception of approximately $4\%$
of exact zero modes, disappear~\footnote{Note that a similar effect was
observed in Ref.~\cite{gattringer} where chirally improved lattice
Dirac operator was studied.}. For the modes remaining in the spectrum
the value of IPR is of order unit. In other words, removal of the
vortices destroys topology-related Dirac modes.

%-----------------------------------------------------------------
%-----------------------------------------------------------------
%-----------------------------------------------------------------

As is mentioned above the change (\ref{modification}) affects a 3d sub-volume
of the whole lattice. Further information on the manifold crucial for the
chirality breaking can be provided by observations on the shrinking
of the localization volume as function of the lattice spacing.
If indeed, the volume shrinks to points, d=0, as favored
by the data above, then it is only a small subspace of the 3d volume
that is related to chiral symmetry. Pure geometrically, such points could well
be the points of self-intersection of the vortices.
Theoretically, possible connection between self-intersections of the vortices and
zero modes was considered, within a particular model, in Ref. \cite{reinhardt}.
In principle, this conjecture on the relation between self-intersections
of the vortices and localization of (near) zero modes could be investigated
through direct observation. At the moment, however, such data are not
available.

\section{Conclusions}

To summarize, we have studied properties of low-lying Dirac modes in the vacuum state
of SU(2) gluodynamics using overlap fermions. We work with the original
gauge fields, which fluctuate on the scale of the lattice spacing, without any smoothening.
The Witten-Veneziano and Banks-Casher relations predict scaling laws for densities
for exact and near zero modes, respectively. Namely, these densities should depend only
on the physical scale, $\Lambda_{QCD}$. From the point of view of the lattice measurements,
these are strong constraints. The results of the measurements do agree with
the theoretical expectations.

A novel feature uncovered by the lattice simulations is that the low-lying
modes are localized on the volumes which shrink as a power of the lattice spacing $a$.
For exact zero modes the power is close to three, $V_{\lambda=0}\sim a^3$
while for near-zero modes it is rather four, $V_{\lambda\to 0}\sim a^4$,
for error bars see Figure~\ref{fig:ipr_vs_a}.

Moreover, the removal of the central vortices from
the original configurations eliminates all the topological fermionic modes.
This observation favors speculations that the lumps of the topological charge are
associated with self-intersections of the vortices.

Note that even if this picture is correct, this does not automatically imply that the
instantonic picture is wrong. There could be two dual pictures of the
same effect, valid for measurements with various resolutions.
In terms of soft fields, instanton-like fluctuations could dominate.
In terms of fields measured with fine resolution, of order $a$,
the fine-tuned surfaces
could be responsible for the lumps of the topological charge.
It goes without saying that the hypothesis on the crucial role of the surfaces is to be
scrutinized much further in the lattice measurements.
If existence of the dual pictures is confirmed by measurements,
one could speculate that this, empirical duality reflects duality
on the fundamental level.
The vortices themselves could be identified  with dual strings \cite{vz},
while their self-intersections could be basic element in the mechanism of the chiral symmetry breaking in
the dual picture.

One cannot rule out that the lattice-spacing
dependence we found will change on finer lattices. In principle, it could happens that
the $\sim a^{-4}$ dependence of IPR shown on Figure~\ref{fig:ipr_vs_a} will eventually flatten
in the limit $a\to 0$ so that the LDEs localization will identify some 4-dimensional objects
of {\it finite} physical extent.  Note however, that the characteristic size of these objects
is definitely less than 0.1 fm and likely to be much smaller than that.
Although this scenario is not excluded {\it a priori}
and requires further investigations, it seems artificial to our mind.

~

The authors are grateful to J.~Greensite, S.~Olejnik, G.~Schierholz, T.~Suzuki
and to the members of ITEP lattice group for stimulating discussions. The invaluable assistance
of G.~Schierholz and T.~Streuer in the overlap operator implementation is kindly acknowledged.
This work was partially supported  by  the EU Integrated Infrastructure Initiative Hadron Physics (I3HP)
grant under contract RII3-CT-2004-506078 and by grants RFBR-05-02-16306a, RFBR-05-02-17642, RFBR-0402-16079,
RFBR-03-02-16941. F.V.G. was partially supported by INTAS YS grant 04-83-3943.
%=================================================================================================

\end{document}